\documentclass[fleqn,usenatbib]{mnras}

\usepackage{newtxtext,newtxmath}
\usepackage[T1]{fontenc}

\DeclareRobustCommand{\VAN}[3]{#2}
\let\VANthebibliography\thebibliography
\def\thebibliography{\DeclareRobustCommand{\VAN}[3]{##3}\VANthebibliography}

\usepackage{graphicx}	% Including figure files
\usepackage{amsmath}	% Advanced maths commands
\newcommand{\kms}{\,km\,s$^{-1}$}

\title[Smith Cloud Filaments]{Properties of 3D HI Filaments in the Smith High Velocity Cloud}

\author[Holm-Hansen et al.]{
Colin Holm-Hansen,$^{1,2}$
M. E. Putman,$^{2}$
D. A. Kim$^{2}$
\\
$^{1}$Department of Physics, Columbia University, New York, NY 10027, USA\\
$^{2}$Department of Astronomy, Columbia University, New York, NY 10027, USA\\
}

\date{Accepted XXX. Received YYY; in original form ZZZ}

\pubyear{2023}

\begin{document}
\label{firstpage}
\pagerange{\pageref{firstpage}--\pageref{lastpage}}
\maketitle

\begin{abstract}
We present findings of 3D filamentary structures in the Smith Cloud, a high-velocity cloud (HVC) located at $l=38^{\circ}$, $b=-13^{\circ}$. These data represent the first detection of velocity-resolved 3D \ion{H}{i} filaments within an HVC. We use data from the Galactic Arecibo L-Band Feed Array \ion{H}{i} (GALFA-\ion{H}{i}) along with our new filament detection algorithm, \texttt{fil3d}, to characterize these structures. In this paper, we also discuss how different input parameters affect the output of \texttt{fil3d}. We study filaments in the local interstellar medium (ISM) and compare them to those found in the Smith Cloud. Based on thermal linewidth estimations we find supporting evidence that the Smith Cloud filaments are part of its warm neutral medium. We also find a relationship between thermal linewidth and the $v_{LSR}$ of the filaments. We study the plane-of-sky magnetic field as traced by Planck 353 GHz polarized dust emission along the line of sight and find none of our filament populations are aligned with this tracer of the magnetic field. This is likely related to their location close to dynamic processes in the Galactic Plane and/or the low column density of the filaments relative to emission in the Plane. The results show that 3D HI filaments are found in a wide range of Galactic environments and form through multiple processes.  
\end{abstract}


%%%%%%%%%%%%%%%%%%%%%%%%%%%%%%%%%%%%%%%%%%%%%%%%%%

%%%%%%%%%%%%%%%%% BODY OF PAPER %%%%%%%%%%%%%%%%%%

\section{Introduction}
\label{intro}

The interstellar medium (ISM) is abundant in complex morphology. It has been shown that there exists an intricate network of molecular filaments that may be linked to star formation \citep{Molinari2010,Arzoumanian2011,André2014}. Recently, \ion{H}{i} filaments have also been found at varying Galactic scales \citep{McClureGriffiths2006,Clark2014, HI4Pi2016, Kalberla2016, Soler2020, Soler2022}. Unlike molecular clouds, these objects are not self-gravitating, meaning they must form due to factors in their environment. Given that they have been detected in several environments at varying scales and densities, multiple formation mechanisms may exist. In order to understand their formation, a plethora of work has been underway studying their physical properties. It has been shown that these filaments tend to align with the ambient magnetic field traced by polarized dust emission. \citep[][]{McClureGriffiths2006,Clark2014, Kim2023}. For molecular filaments, a perpendicular magnetic field alignment is thought to be preferred \citep{Federrath2016,Gomez2018,Hennebelle2019}.

Characterizations of HI filaments in regions beyond the Galactic ISM are beginning to take shape. \cite{Ma2023} studied filamentary structure in regions of the Small Magellanic Cloud and found possible evidence of alignment with its magnetic field. \cite{Jung2023} have shown in MHD simulations that high-velocity clouds (HVCs) draped in a magnetic field should have filaments along their direction of motion. However, up until now, the quantification of observed \ion{H}{i} filaments in a HVC has not been done. 

The Smith Cloud \citep{Smith1963} is one of the most thoroughly studied HVCs with a known distance of 12.4 kpc \citep{Putman2003, Lockman2008, Wakker2008}. Like many other HVCs it has a comet-like morphology with a compact head and more diffuse tail. Its head is located at approximately $l=38^{\circ},b=-13^{\circ}$, and the whole structure extends approximately 1 $\times$ 3 kpc across the sky. Mass estimates suggest the Smith Cloud has approximately $10^6 M_{\sun}$ in both neutral \ion{H}{i} and \ion{H}{ii} \citep{Lockman2008,Hill2009}. Its magnetic field strength has been mapped using Faraday Rotation \citep{Hill2013,Betti2019}. \cite{Lockman2008} computes a z-height for the Smith Cloud of -2.9 $\pm~0.3$ kpc, which we adopt for this paper. The kinematic analysis done by \cite{Lockman2008} has shown that it is on a trajectory toward the Galactic plane, and will reach it in about 30 Myr.

The origin of the Smith Cloud is still unknown. It has been proposed as having an extragalactic origin \citep{BlandHawthorn1998} and also a local origin as a conglomerate ejected from the disc of the Milky Way \citep{Sofue2004, Marasco2017}. If it is a failed dwarf galaxy it likely already passed through the disc approximately 70 Myr ago and in order for it to have survived this encounter it may need a dark matter halo \citep{Nichols2009}. If it has a local origin, \cite{Alig2018} have shown that the next encounter with the disc in 30 Myr may trigger a burst in star formation. Further mapping the substructure of the cloud can help to better understand its origin and fate through comparing the observed structure to simulations of halo clouds growing and deteriorating as they interact with their surrounding medium (Porter et al. in prep.).  

The Smith Cloud substructure may also provide insight into the orientation of its magnetic field. Simulations indicate the Smith Cloud could not have survived for as long as it has in the Galactic Halo without some sort of stabilizing mechanism \citep{Nichols2009, Alig2018}. One stabilizing candidate is the cloud's magnetic field \citep{McClure-Griffiths2010}. Due to the discovery that filaments in the local \ion{H}{i} gas align with the plane-of-sky Galactic magnetic field at high latitudes, its possible that filaments in an HVC could also preferentially align with its magnetic field.

In this paper, we investigate the structure and properties of 3D \ion{H}{i} filaments in the Smith Cloud. We compare their properties to filaments at lower velocities along the same line of sight. Section \ref{data} introduces the data and reviews the procedures for both extracting and analyzing 3D \ion{H}{i} filaments using our 3D filament finding algorithm, \texttt{fil3d}. Section \ref{fil3dparams} discusses the effects of \texttt{fil3d} parameters on the final filament populations and goes over the selections made for this study. Section \ref{results} shows the results of running \texttt{fil3d} on these data and investigates filament properties such as physical characteristics and magnetic field orientation. We discuss these results and present our conclusions in Sections \ref{discussion} and \ref{conclusion}. 

\section{Data \& Methods}
\label{data}
\begin{figure}
 \includegraphics[width=\columnwidth]{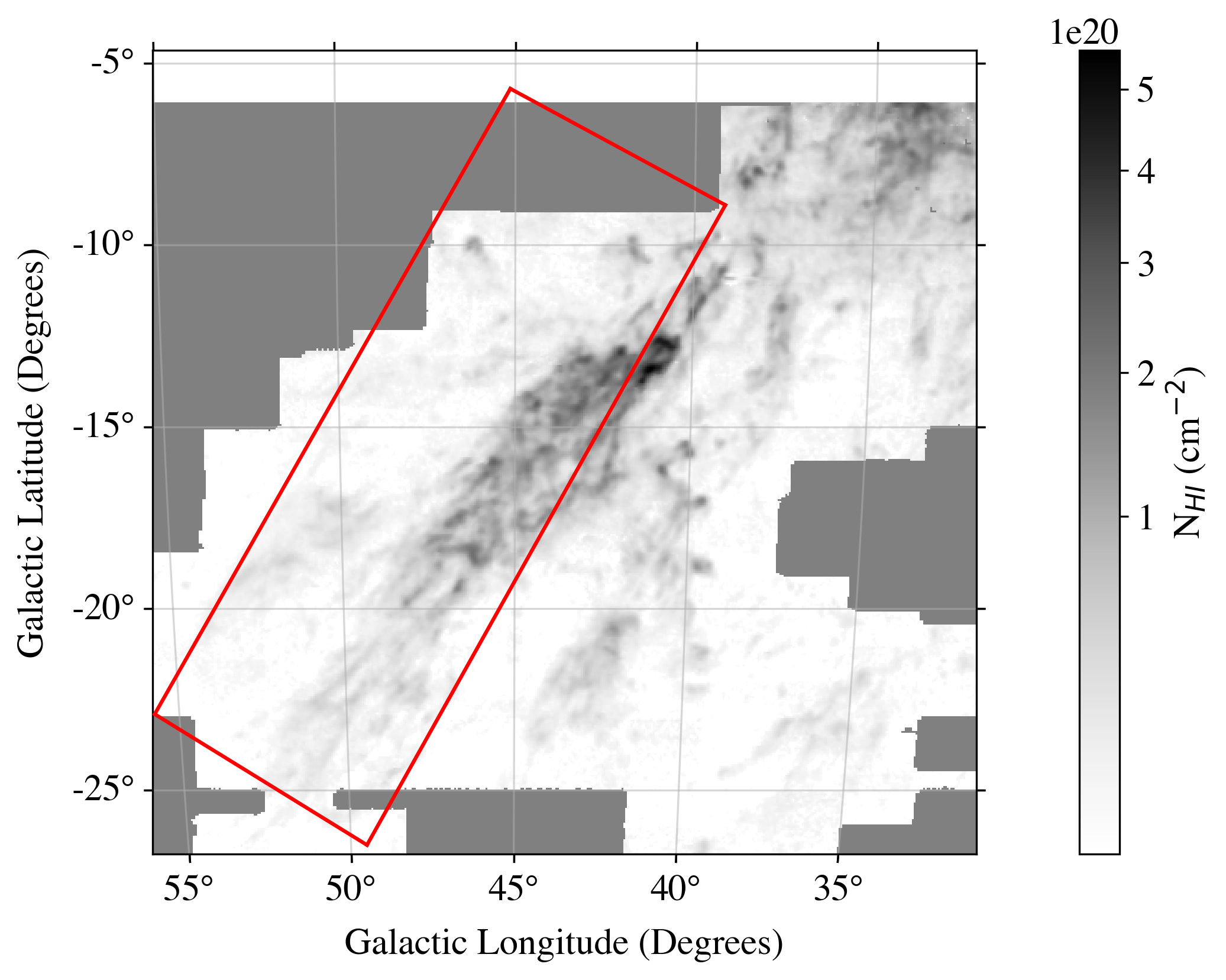}
 \caption{Moment zero map of the Smith Cloud using data from \protect\cite{Lockman2008} integrated from 70-130\kms, with the column density from the moment 0 map shown with the right scale bar. The GALFA-\ion{H}{i} region, which is used for the rest of the paper due to its greater spatial resolution, is shown in a rectangle outlined in red. }
 \label{fig:lockmanregion}
\end{figure}

\begin{figure*}
 \includegraphics[width=.99\textwidth]{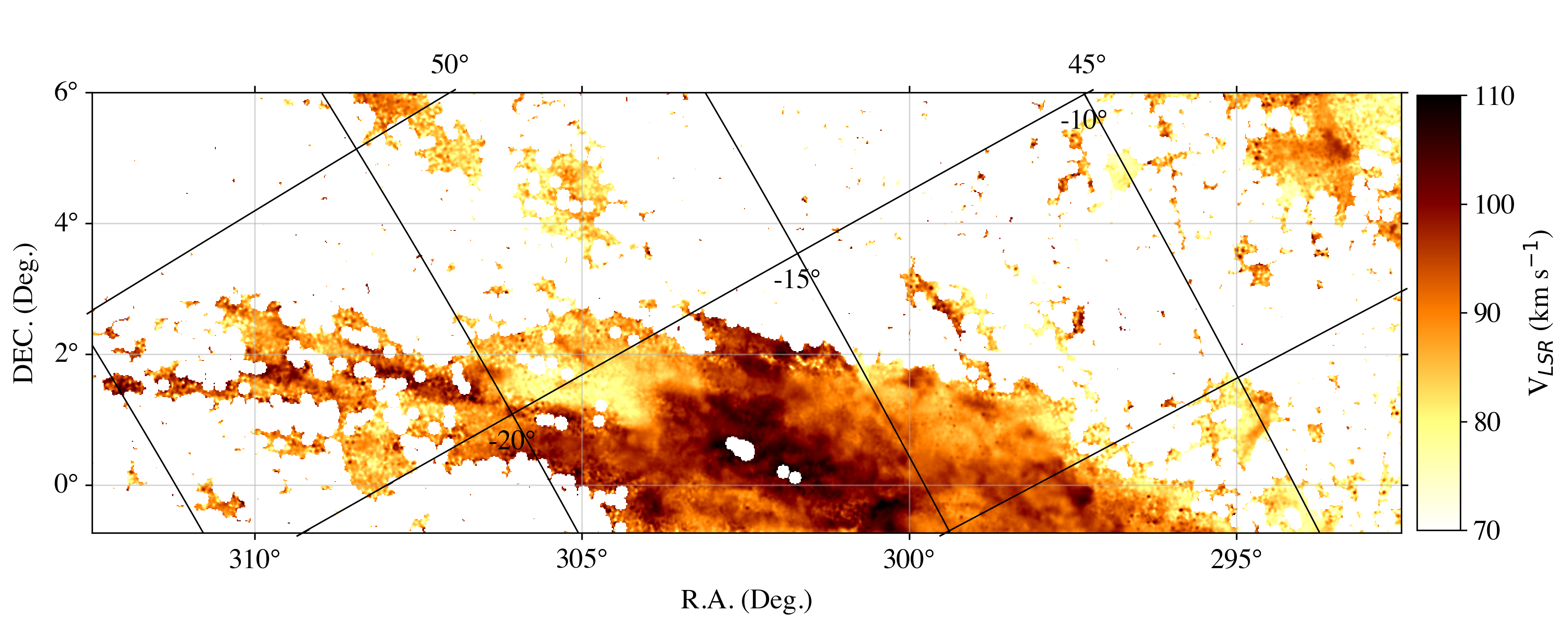}
 \includegraphics[width=.99\textwidth]{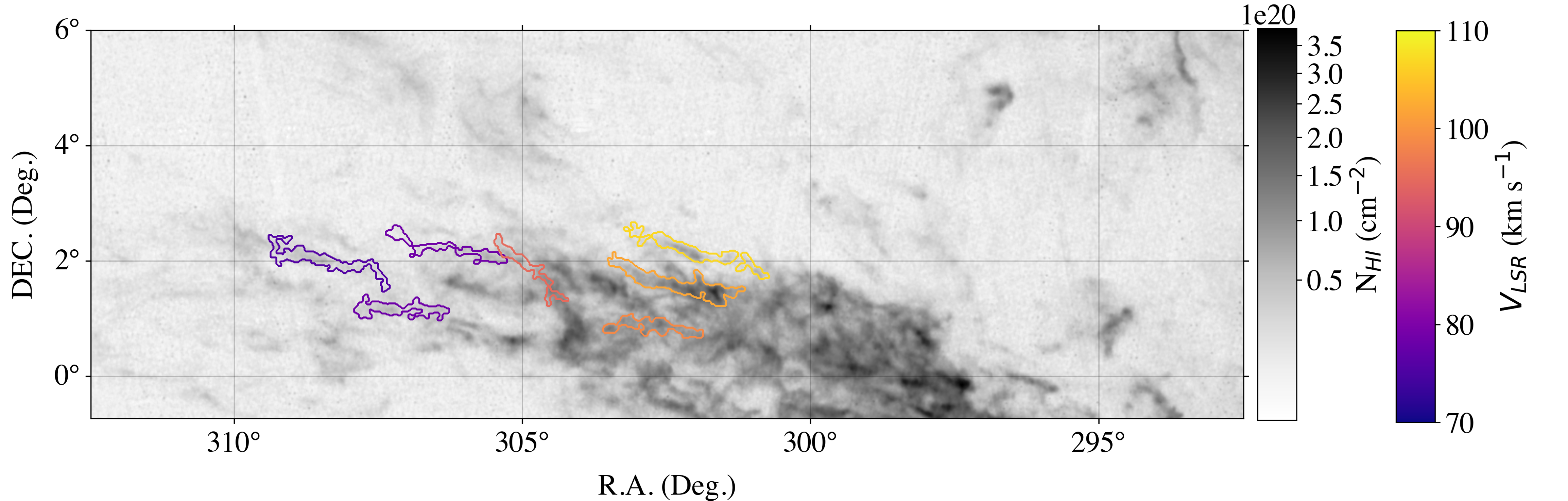}
 \centering
 \caption{Top: Average velocity moment 1 map of the GALFA-\ion{H}{i} region of the Smith Cloud over the same velocity range as Figure \protect{\ref{fig:lockmanregion}} with Galactic coordinates overlaid. The circles in the data are artefacts from the clipping procedure used in the generation of the map and are not physical. Bottom: HVC filament masks overlaid on N$_\ion{H}{i}$ map of the Smith Cloud integrated from 70-130\kms. More information regarding the data can be found in \protect\cite{Peek2018}.}
 \label{fig:upperfilaments}
\end{figure*}

We extract filaments with data from the Galactic Arecibo L-Band Feed Array \ion{H}{i} (GALFA-\ion{H}{i}) survey \citep{Peek2018}. GALFA-\ion{H}{i} is highly effective at locating filaments due to its sensitivity of 150 mK per 1\kms channel and spatial resolution of 4\arcmin. The data are gridded to 1\arcmin x 1\arcmin pixels. We downloaded a custom cube from the GALFA-\ion{H}{i} website \footnote{https://purcell.ssl.berkeley.edu/}, using the 0.739\kms channel spacing for spectral resolution over a total velocity range of -68\kms < $~v_{LSR}~$ < 188\kms. We opt for the 0.739\kms channels over the narrow 0.184\kms channels to account for the greater line widths and lower column densities of HVCs. The velocity-position slice of the cloud in \cite{Lockman2008} indicates an approximate $v_{LSR}$ range of $60-120$\kms for the cloud, meaning our range is sufficient to capture the entire extent of the cloud as well as the Galactic co-rotating gas around 0\kms. The total region of sky sampled in Right Ascension ($\alpha$) and Declination ($\delta$) is 292\degr < $\alpha$ < 312\degr and -1\degr < $\delta$ < 6\degr. The trade-off with using the high spatial resolution GALFA-\ion{H}{i} data over other survey data is that we are located at the minimum declination of GALFA-\ion{H}{i}'s observation region (-1\degr17\arcmin), whereas the main component of Smith Cloud extends as far as -2\degr \citep[see][]{Lockman2008}.  
Figures \ref{fig:lockmanregion} and \ref{fig:upperfilaments} show the Smith Cloud as shown in the Robert C. Byrd Green Bank Telescope (GBT) data \citep{Lockman2008} and the GALFA-\ion{H}{i} data respectively. For magnetic field information, we employ the \textit{Planck} 353 GHz Stokes linear polarization maps \citep{Planck2020}. We smooth the resolution of the \textit{Planck} maps to a full width at half-maximum (FWHM) of 1\degr~from their original resolution of 4.9\arcmin in order to improve the maps' signal-to-noise ratio.

We use the \texttt{fil3d} algorithm to identify filaments in GALFA-\ion{H}{i} as discussed in \cite{Kim2023} and Putman et al. (in prep.). In an attempt to capture the entirety of the Smith Cloud, we also tried running \texttt{fil3d} on data from the GBT from \cite{Lockman2008} as well as from the \ion{H}{i} 4$\pi$ survey \citep{HI4Pi2016} using a wide range of settings, but found with the larger beam widths (10\arcmin and 16\arcmin, respectively), filaments are not resolved enough to be captured by \texttt{fil3d} (although see Figure \ref{fig:galfafilfinder} for the 2D filament map using the GBT data). We follow the same procedure for extracting filaments laid out in \cite{Kim2023} and Putman et al. (in prep), which can be consulted for more details. We begin by applying an unsharp mask (USM) with a 30\arcmin Gaussian beam to each channel in the spectral cube, which acts as a high-pass filter, removing diffuse emission. Following this we run \texttt{FilFinder} \citep{Koch2015} to identify 2D filamentary structures in each channel. Each 2D filament is referred to as a node. If any two nodes in adjacent channels share 85\% of pixels in common, they are considered to be part of one composite 3D object, known as a tree. Once a tree is formed, it is finalized by repeating this process until adjacent channels have no significant overlap. The final mask for a tree is the aggregated mask of each of its nodes, called the merged mask.

\begin{table}
\centering
 \caption{Input parameters used for this study. The first four variables with underscores are \texttt{Filfinder} parameters. Aspect ratio is a filtering parameter,and the remaining ones are inputs to \texttt{fil3d}.}
 
 \label{inputparams}
 \begin{tabular}{lc}
  \hline
  Name & Value \\
  \hline
  \texttt{scale\_width} & 0.1 pc \\
  \texttt{smooth\_size} & 0.05 pc \\
  \texttt{adaptive\_threshold} & 0.2 pc \\
  \texttt{size\_threshold} & 0.12 pc$^2$ \\
  Aspect Ratio & 1:6 \\
  Overlap Threshold & 85\% \\
  USM Kernel & 30\arcmin \\
  \hline
 \end{tabular}
\end{table}

After generating trees, we validate them based on several criteria. First, only trees with an aspect ratio of 1:6 or greater are accepted to ensure filament-like morphology (discussed further in \ref{aspect ratio}). Since \texttt{FilFinder} tends to pick up more noise at the edges, any trees with a merged mask along the edge of the data are also discarded. If filaments pass these first two criteria we inspect their median intensity within the merged masked area as a function of their velocity (see \cite{Kim2023}, Figure 4). This is done since most \texttt{fil3d} trees only span 2-3 channels. As such, we consider the FWHM of the Gaussian profile to be a more physical line width for the filaments as opposed to only the \texttt{fil3d} detected channels. To ensure the validity of this assumption, we require that the median velocity channel detected by \texttt{fil3d} falls within one standard deviation of the best fit Gaussian profile. Finally, we visually inspect the merged mask contours overlaid on a moment zero map of the USM data integrated over the FWHM of their Gaussian profile. Since we are interpreting this range as the true linewidth of a filament, we require that the moment zero map integrated over this range be consistent with the original \texttt{fil3d} merged mask. Therefore, only filaments whose merged masks match the data over their FWHM range are accepted. Filaments that pass all of these criteria are then saved in the final set. Applying this to our dataset we find an initial return of 135, which becomes 72 after the filtering process. The final batch covers a velocity range (taken to be the velocity of the median channel of the filament found by \texttt{fil3d}) of -6\kms < v$_{LSR}$ < 107\kms.

For the plane of sky magnetic field data, we follow the same procedure as \cite{Kim2023}, using the supplied Stokes data to obtain the dust polarization $\psi$ in the Galactic IAU convention.

Since polarized dust grains align perpendicularly with the ambient magnetic field, we define the plane-of-sky magnetic field orientation $\phi$ at a given location to be:

\begin{equation}
    \phi = \frac{1}{2}\arctan(U,-Q),
\end{equation}

where $U$ and $Q$ are the normal Stokes parameters. We note that the angle $\phi $ is a 90\degr rotation from $\psi$. From this, we compute the average B-field orientation value $\overline{\phi}$ in the area of each filament:
\begin{equation}
    \overline{\phi} = \frac{\sum_{j}^{}\sum_{i}^{} \phi(x_{i},y_{j})}{A}
\end{equation}
where A is the area of the merged mask, and $\phi(x_{i},y_{j})$ is the magnetic field orientation at location $(x_{i},y_{j})$ within the area of the merged mask. The orientations of the filaments themselves, denoted as $\theta$, are computed using the Rolling Hough Transform (RHT) as described in \cite{Clark2014}. We define the difference between these two values, $|\theta - \overline{\phi}|$, to be the mean magnetic alignment of a filament.
We also compute the polarization fraction $p$ for each filament:
\begin{equation}
    p = \frac{\sqrt{U^2 + Q^2}}{I}
\end{equation}
which we use to help discuss our findings in Section \ref{discussion}.

\begin{figure*}
 \includegraphics[width=0.99\textwidth]{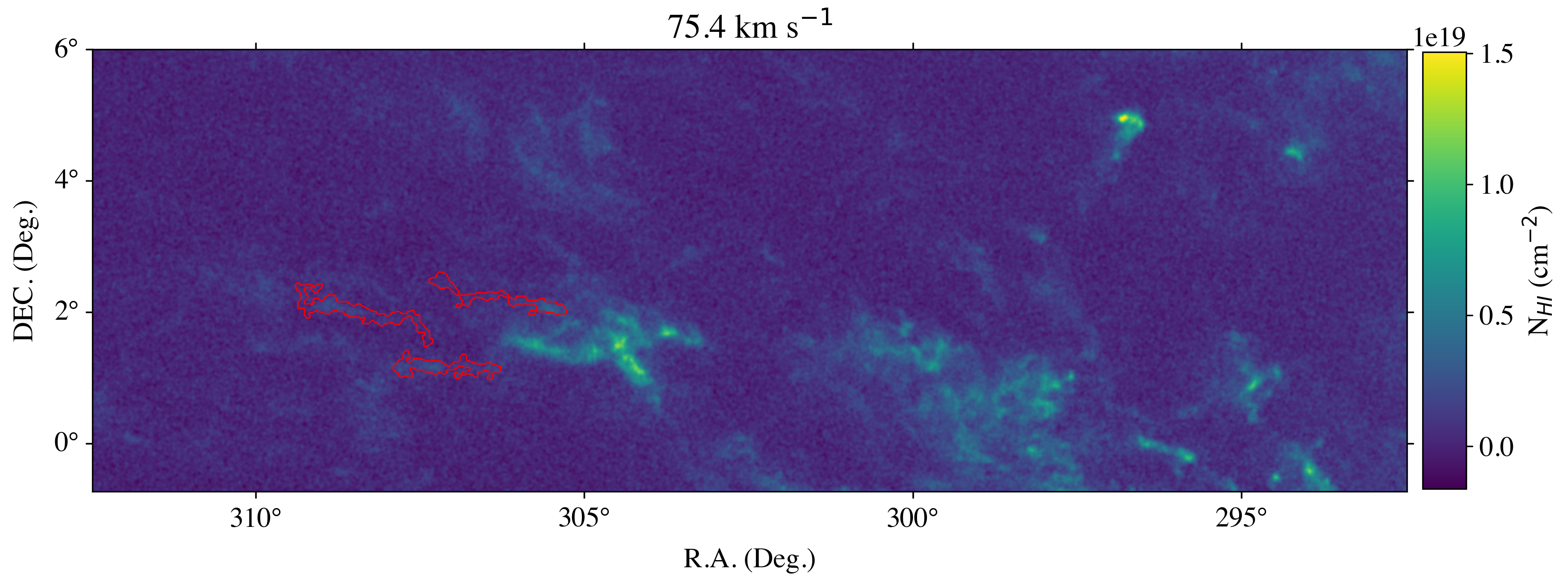}
 \includegraphics[width=0.99\textwidth]{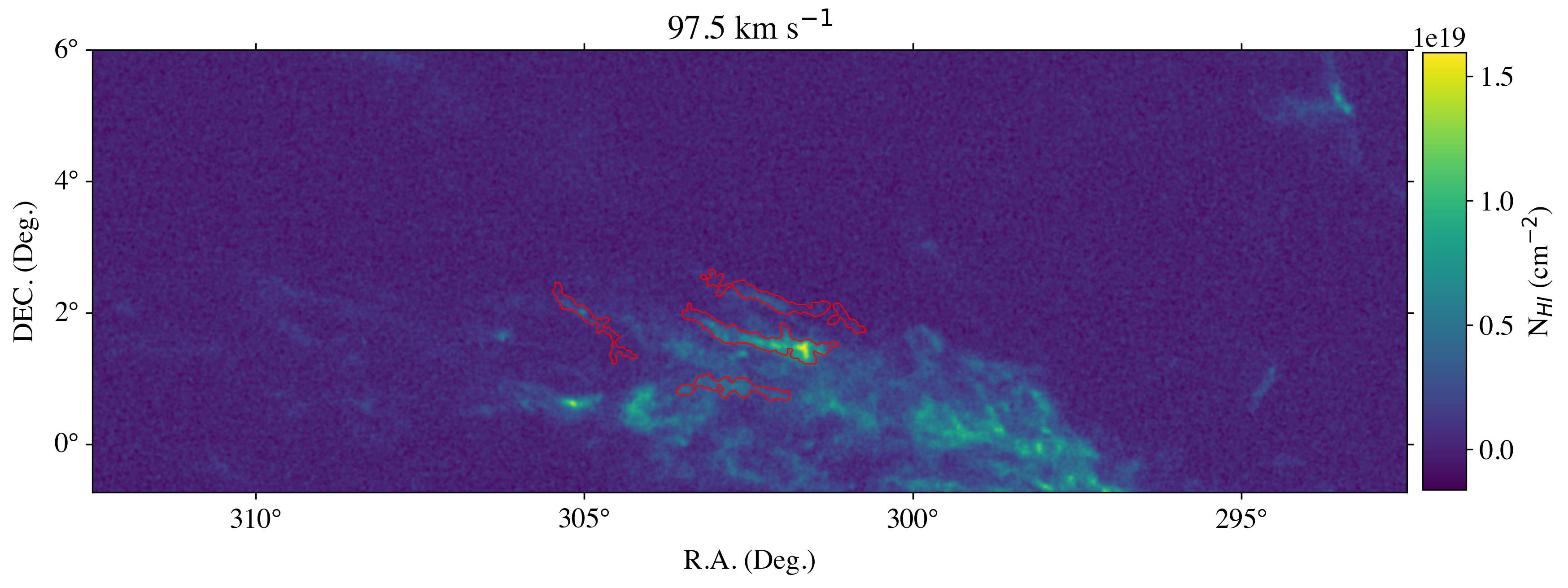}
 \caption{HVC filaments overlaid on individual channels: the upper panel shows the filaments in the tail over a channel with a velocity of 75.4\kms, and the body filaments are shown over a channel at 97.5\kms. The channels were chosen to approximately correspond to the median velocity of the filaments shown, as depicted in Figure \ref{fig:upperfilaments}.}
 \label{fig:tailandbodychannels}
\end{figure*}

\section{\textit{fil3d} Dependencies}
\label{fil3dparams}
The output generated from \texttt{fil3d} is sensitive to its multiple input parameters, including inputs made to \texttt{Filfinder}, the overlap percentage, the aspect ratio filter, and the width of the USM Gaussian kernel. Different inputs should be expected to work better for different data sets as well as in different regions. Currently, these parameters need to be manually fine tuned to best accompany one's dataset (see Section \ref{conclusion} for further discussion on this topic). Overall, we find that tweaking these parameters often changes the total number of filaments, but the overall trends and findings discussed in Section \ref{results} stay the same. Here we discuss the selections made for this study. The general goal of fine tuning the parameters is to ensure that visibly apparent filaments are captured with relatively accurate properties, as well as to minimize the appearance of artefacts in the final batch of data. We define an artefact to be anything that upon visual inspection is deemed to be non-physical, such as a 90$\degr$ corner, a perfectly straight edge, or a node constructed around noise in the data. For these selections, we find the multiple steps of validation criteria to serve as a means to minimize artefacts in the final population. For example, lowering the size threshold or the minimum aspect ratio requirement inherently leads to more artefacts being included in the initial return sample. However, we find that when these changes are made, the percentage of filaments removed by the velocity filter increases significantly. As such, the multiple criteria help ensure the validity of the final population. The full list of parameters are shown in Table \ref{inputparams}. 

\begin{figure}
 \includegraphics[width=\columnwidth]{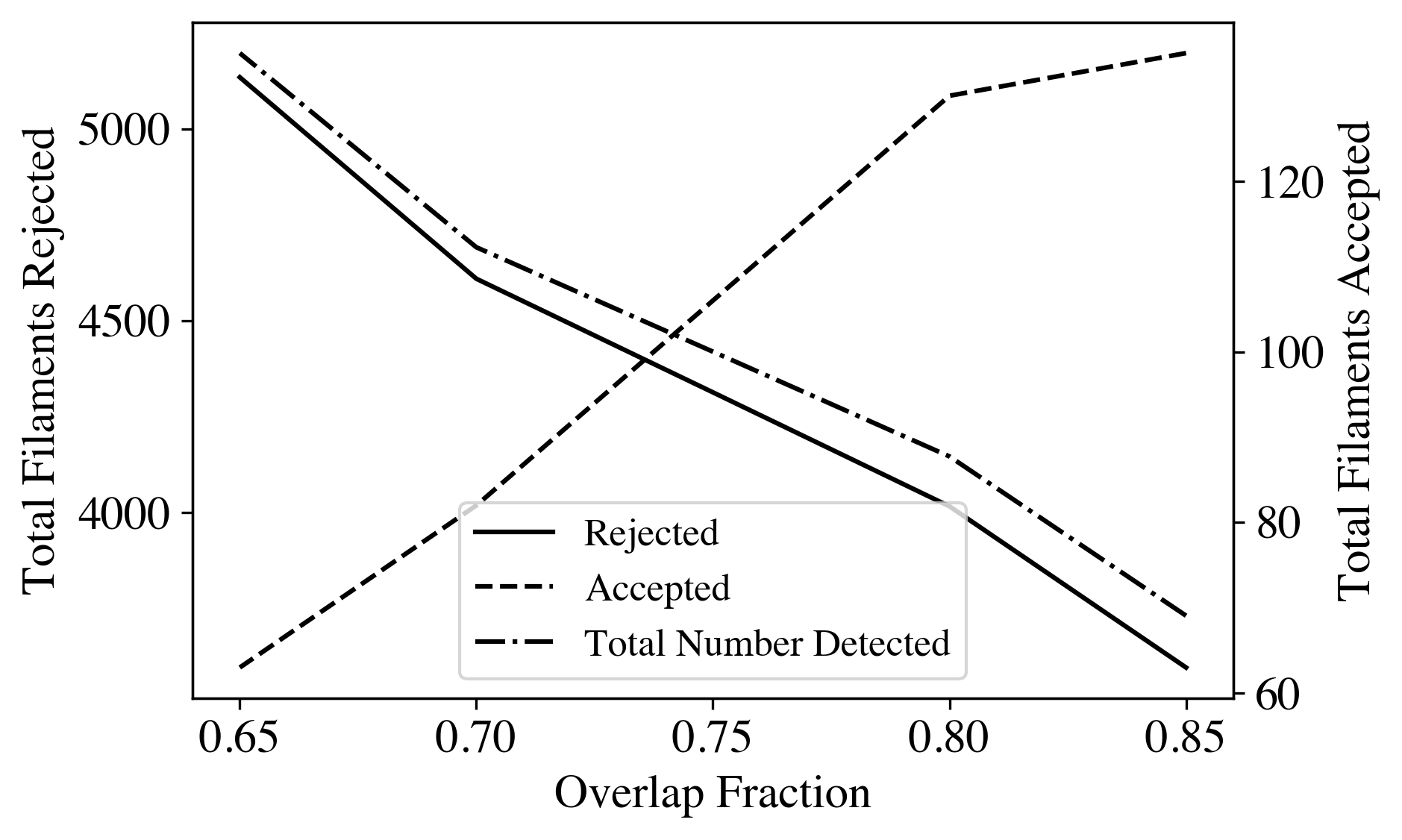}
 \caption{Comparison of how the filtering process changes as a function of the chosen overlap fraction. The solid line shows the total number of detected filaments before any filtering is done. The solid line shows the number of filaments rejected by an aspect ratio filter of 1:6 compared to the corresponding overlap fraction used for filament detection in \texttt{fil3d}. A lower overlap fraction causes rounder filaments on average, increasing the number of filaments excluded from the final set. The dashed line shows the total number of filaments satisfying at least a 1:6 aspect ratio as a function of the overlap fraction.
 }
 \label{fig:aspectratio}
\end{figure}

\subsection{\textit{FilFinder} Parameters}
The nodes drawn from each channel are partially dependent on the parameters used for \texttt{Filfinder}. For these, we start by defining a characteristic \texttt{scale\_width} and defining the rest of our variables with respect to it. Doing so helps reduce the number of free parameters when running the algorithm. The scale width directly influences the target filament size for the algorithm to pick out. For this value, we found 0.1 pc to work the best. 0.15 pc and 0.05 pc were also tried, but with the former, the number of filaments drops drastically and the latter picks up too many artefacts. Other inputs to \texttt{Filfinder} include \texttt{smooth\_size} which was set to $0.5*$\texttt{scale\_width}, and \texttt{adaptive\_threshold} which was set to $2*$\texttt{scale\_width}. \texttt{smooth\_size} determines the magnitude of the beam applied to the image to remove small noise variations, and \texttt{adaptive\_threshold} is a parameter related to the expected filament size. \texttt{border\_masking}, which excludes the edges from the search region, was set to false. More details on these parameters can be found in the \texttt{Filfinder} documentation. 

The \texttt{Filfinder} parameter that was given the most consideration was \texttt{size\_threshold}, which is the minimum size required for an individual node to be considered real. We experimented with using both very large and very small values for this parameter. We started with $8*$\texttt{(scale\_width}*2)$^2$ (used in \cite{Kim2023} and Putman et al.), but found that this left out visible filamentary structure in the Smith Cloud. We then tried dropping it down to 5 pix$^{2}$ with the thought that any artefacts picked up would be filtered out through our other validation criteria. However, we found that under this setting many filaments ended up with nodes only a few pixels in size, and some filaments recovered at higher size thresholds had additional nodes at the beginning or end of their channels that were squares 5-10 pixels in size. We eventually settled on $3*$\texttt{(scale\_width}*2)$^2$ as a compromise between the two extremes. The smallest nodes were inspected visually at this threshold and none appeared to be artefacts. 
\subsection{Aspect Ratio} \label{aspect ratio}

\begin{figure*}
 \begin{tabular}{ll}
 \includegraphics[width=0.6\textwidth]{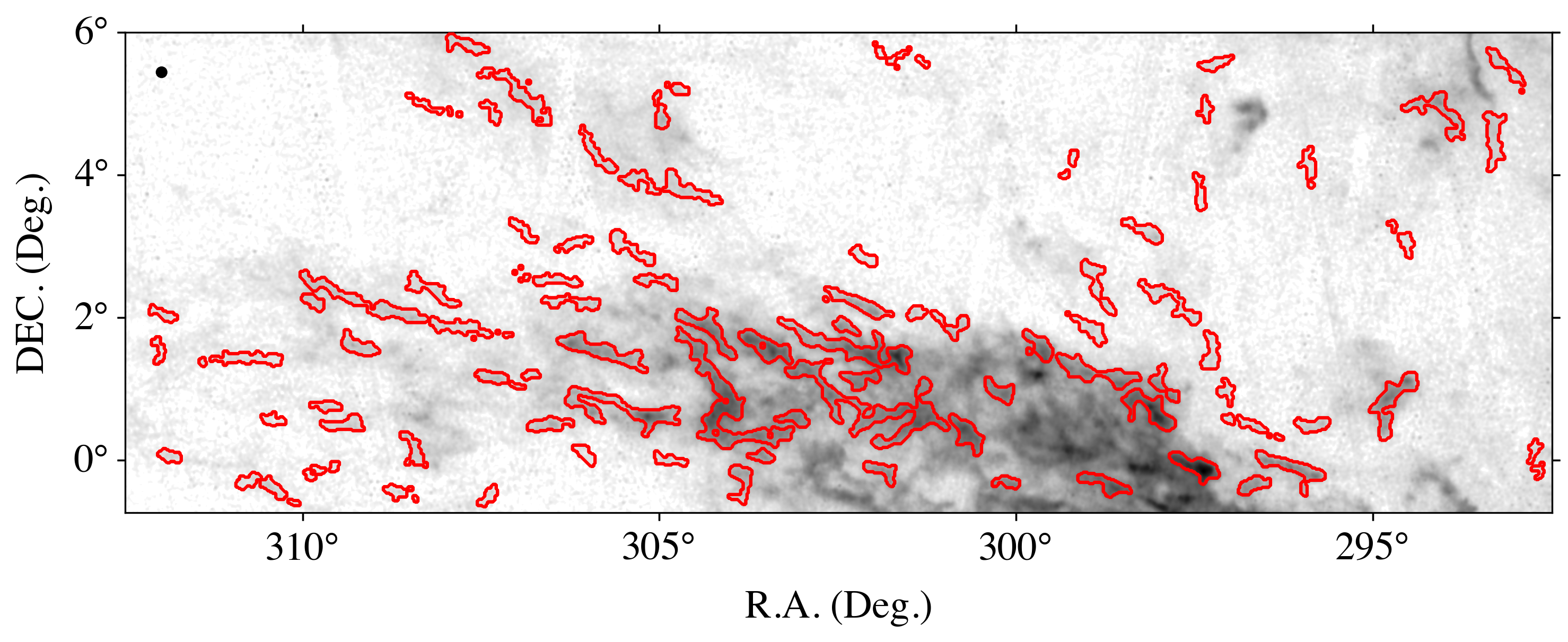}
 &
 \includegraphics[width=0.65\columnwidth]{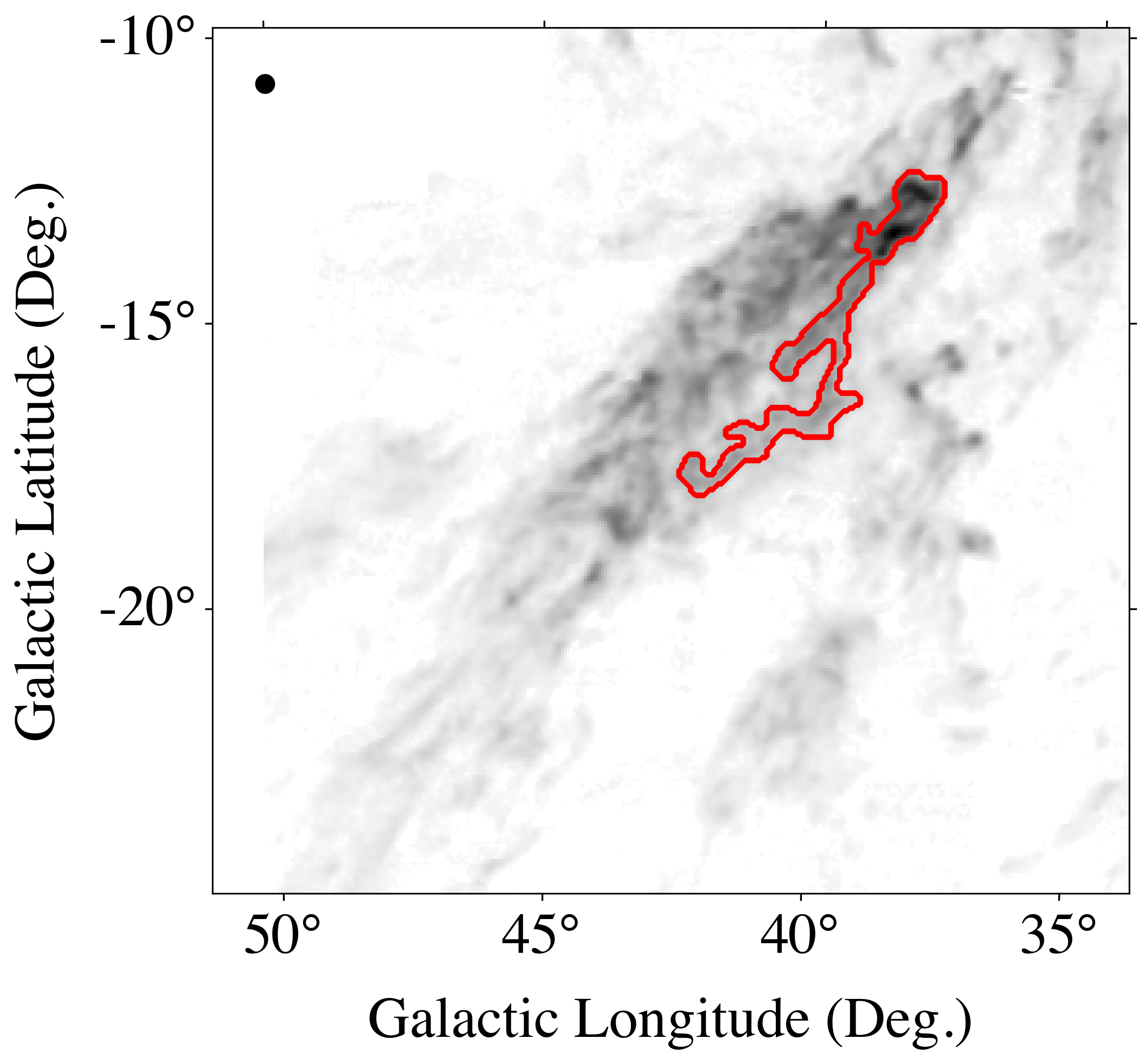}
  \end{tabular}
 \caption{Left: \texttt{Filfinder} results run on the moment 0 map of the GALFA-\protect{\ion{H}{i}} from 70-130\kms region. The beam size is drawn as a black circle in the upper left corner. Right: Filfinder results ran on the moment 0 map in Figure \ref{fig:lockmanregion} with a 1:2 aspect ratio cut, minimum area of twice the beam size (shown as a black circle in the upper left hand corner), and edge filtering applied. The prominent structure centered at (l,b) $\approx$ (40\degr,-15\degr) is outside of the GALFA-\protect{\ion{H}{i}} data.}
 \label{fig:galfafilfinder}

\end{figure*}

We tested several different aspect ratio filters: 1:6, 1:5, 1:4, and even 1:2 was considered to better understand the overall effect on the final filament count. As expected, as the aspect ratio filter drops, the total number of filaments increases. For these data, 1:6 usually returned just under one hundred filaments (For this subsection, the totals referred to do not include any other sort of filtering as discussed in Section \ref{data} other than the aspect ratio filter). 1:5 and 1:4 returned several hundred filaments each, and 1:2 returned over one thousand filaments. As with other parameters, these aspect ratio outputs were inspected visually to determine which produced the most filaments without producing too many artefacts. We selected 1:6 for this study, consistent with what was used in \cite{Kim2023}.  The aspect ratio definition is changed in Putman et. al (in prep) to represent the long axis of the filaments in various regions of the sky more accurately, but this change does not affect the filaments here.

\subsection{Overlap Threshold}
Changes to the overlap threshold between filament nodes had a large effect in this data cube with our aspect ratio cuts.  Somewhat counter-intuitively, the total number of filaments found decreases if the overlap threshold is lowered too far below 85\%. This occurs because as the overlap threshold is lowered, more and more nodes are accumulated into the merged mask structure. This causes filaments to span across more channels on average, but it also causes them to become wider, or to pick up adjacent material that is not associated with them. As such their aspect ratios decrease, causing more filaments to be rejected by the aspect ratio filter. Figure \ref{fig:aspectratio} shows how the total number of filaments rejected by the aspect ratio filter scales with the overlap fraction. There are 5135 filaments rejected using an overlap fraction of 65$\%$, compared to 3594 rejected with an overlap fraction of 85$\%$. From the Figure, it is evident that the quantity of filaments rejected is much greater than the new filaments gained as the overlap fraction is changed, causing the final filament total to actually decrease at a lower overlap fraction.

\section{Results}
\label{results}

\begin{figure}
 \includegraphics[width=\columnwidth]{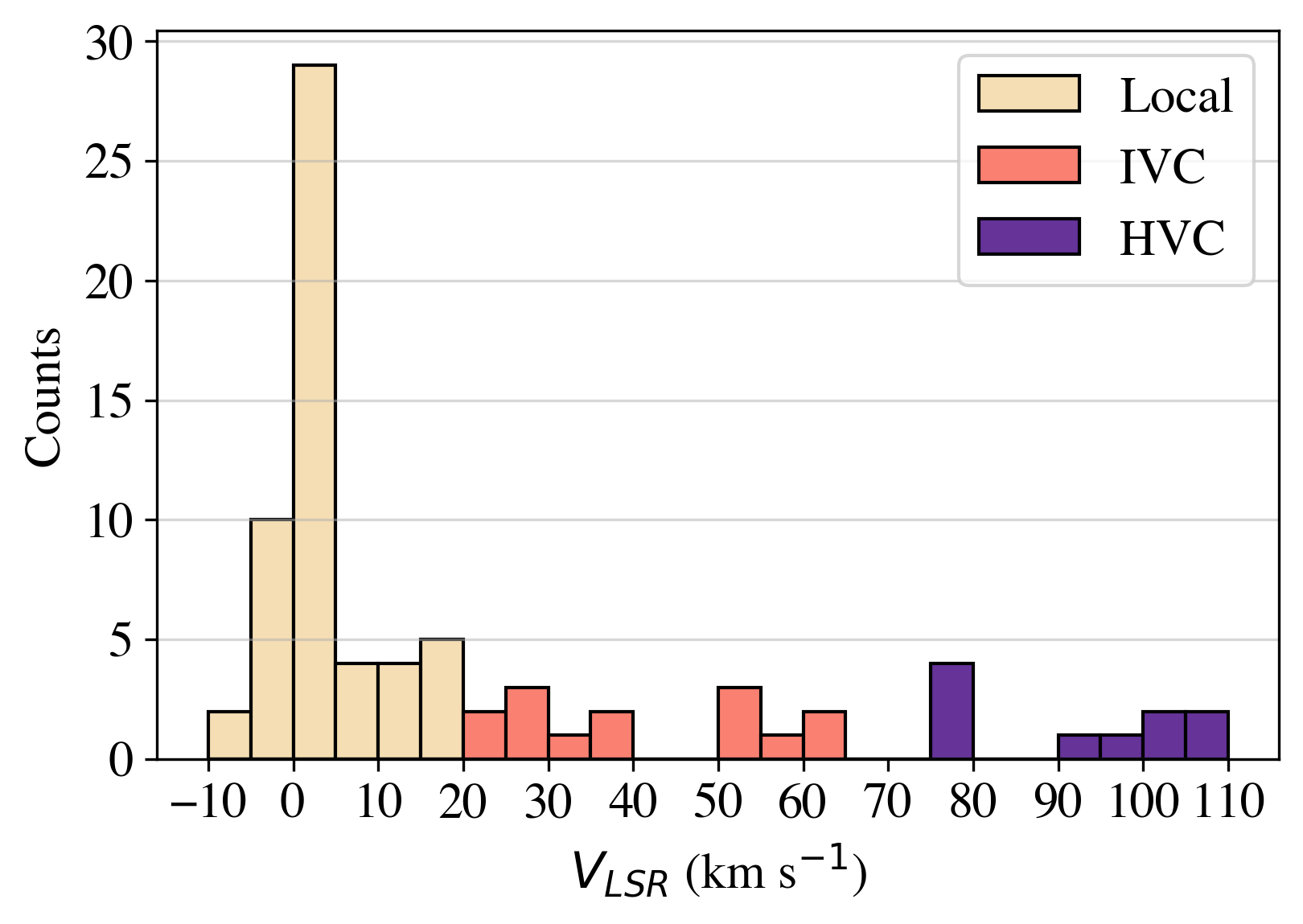}
 \caption{Histogram of the final sample of 3D filaments as a function of their median \texttt{fil3d} $v_{LSR}$. The bins are colored by their grouping (local, IVC, HVC) as described in the text in Section \ref{results}. Each bin has a width of 5\kms.}
 \label{fig:velodistribution}
\end{figure}

\begin{figure*}
 \includegraphics[width=0.99\textwidth]{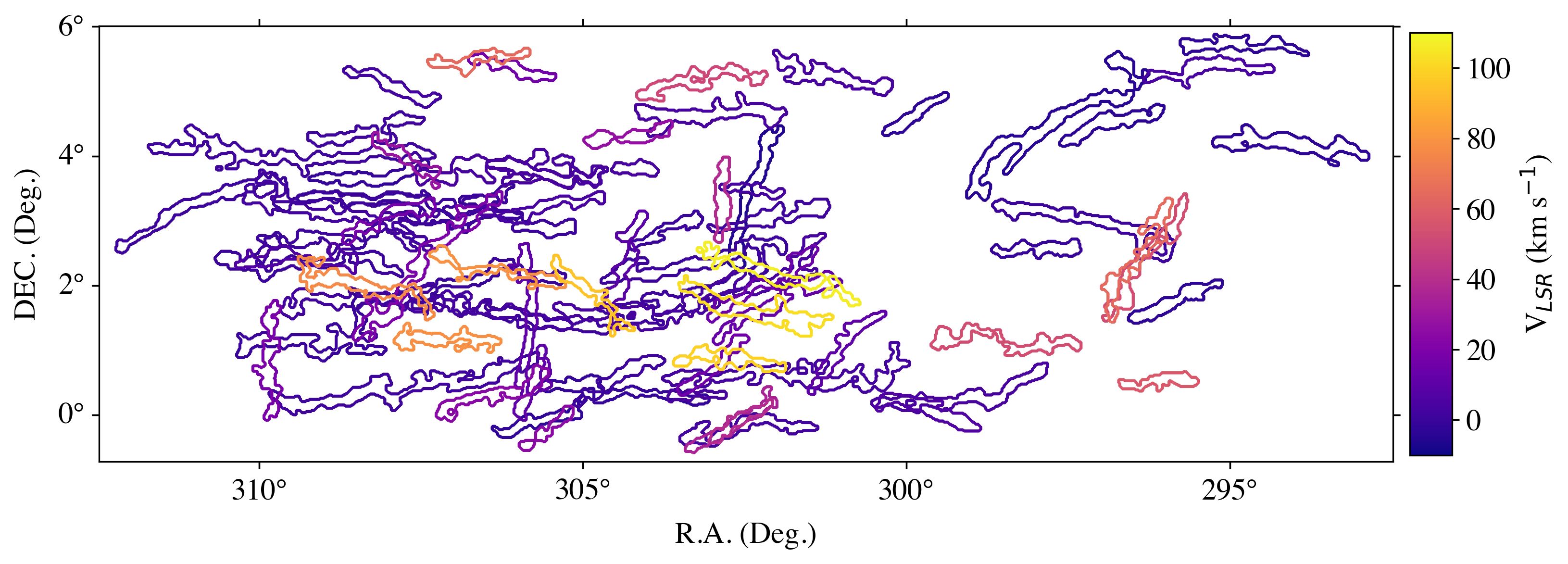}
 \includegraphics[width=0.99\textwidth]{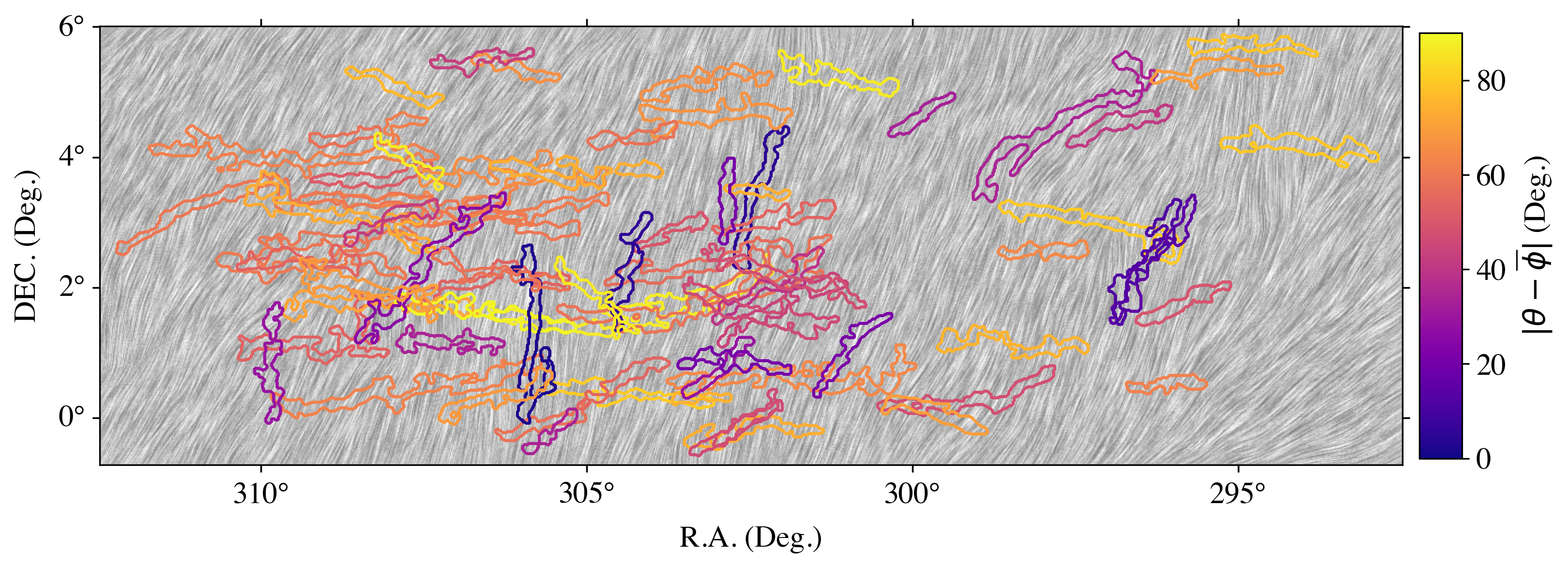}
 \caption{Top: Catalog of all filaments detected. Color indicates $v_{LSR}$ of the median channel of the filament as found by \texttt{fil3d}. Bottom: Magnetic field orientation of all filaments. The filament color shows the computed magnetic field alignment statistic $|\theta -  \overline{\phi}|$, with 0 degrees indicating perfect alignment to the plane-of-sky magnetic field and 90 degrees indicating orthogonal alignment. The background shows magnetic field orientation drawn with line integral convolution. Local filaments appear anti-aligned in this region.}
 \label{fig:localbfield}
\end{figure*}

\begin{figure}
 \includegraphics[width=\columnwidth]{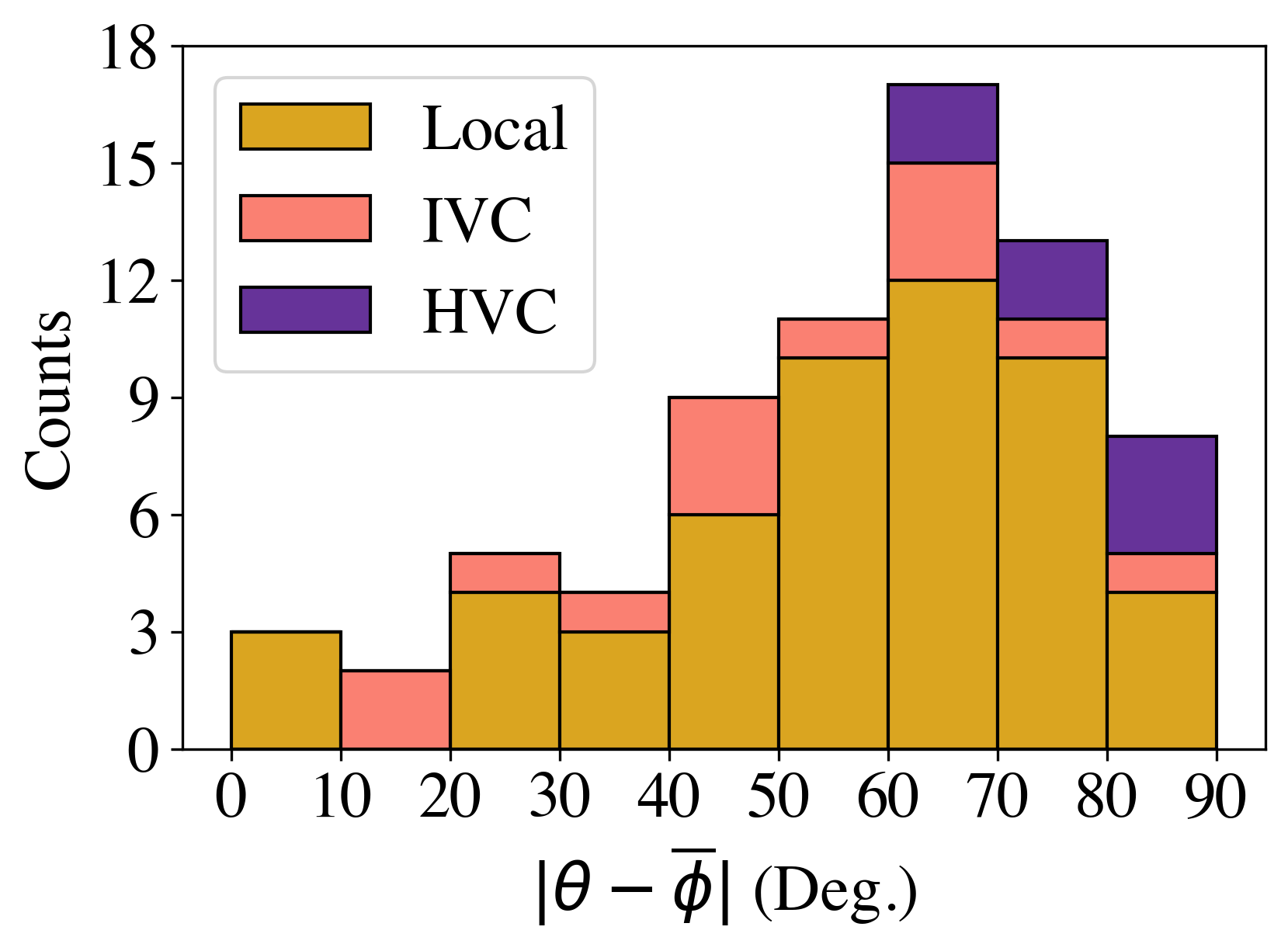}
 \caption{Histogram of filament alignment with the plane of sky magnetic field with a bin size of 10 degrees. Filaments are grouped by region: local (gold), IVC (orange), and HVC (purple). The local filaments are preferentially anti-aligned with the magnetic field, unlike filaments found at a high Galactic latitude \citep{Kim2023}. The plane-of-sky magnetic field inferred from dust polarization is unlikely to capture magnetic field information from the Smith Cloud at a distance of 12.3 kpc.}
 \label{fig:bhist}
\end{figure}
\subsection{Filament Groupings \& Characteristics}

We find a total of 72 3D filaments passing all criteria outlined in Section \ref{data} within the GALFA-\ion{H}{i} data. We note that there is no uniform definition of filaments in the literature, and for our purposes we simply define them to be compact, visibly linear structures which satisfy the filtering criteria outlined in Section \ref{data}. 7 of those filaments are contained within the Smith Cloud velocity range and are shown over the Smith Cloud moment 0 map in Figure \ref{fig:upperfilaments}.  Looking at the distribution of median velocities for each filament as shown in Figure \ref{fig:velodistribution}, we categorize the filaments into three distinct groups. The bulk of filaments with median velocities less than 20\kms are considered to be relatively local structures in the Galactic disc. The second group is a smaller population between 20 and 70\kms, which are largely thought to represent structures in intermediate velocity clouds (IVCs). The remaining group, all with velocities of 70\kms or greater, are the filaments located in the Smith Cloud. This range of velocities for the Smith Cloud is based on Figure 3 of \cite{Lockman2008}, which suggests that prominent Smith Cloud flux ranges from 70-130\kms. The HVC filaments are plotted in Figure \ref{fig:upperfilaments}. The entire group of filaments are plotted in Figure \ref{fig:localbfield}.The HVC filaments plotted over individual Smith Cloud channels is shown in Figure \ref{fig:tailandbodychannels}.

The majority of filaments detected in the Smith Cloud itself are located away from the head, closer to the tail and midsection. Looking at the right panel of Figure \ref{fig:galfafilfinder}, it seems probable there are more filaments in the midsection that are not being detected due to them being partially or completely cut off from the GALFA-\ion{H}{i} region. The filaments that are detected are aligned with the morphology of the Smith Cloud itself. 
There are initially 9 detected after filtering in this region; however, there was a significant overlap in the filament contours for two filaments in the tail and two in the body. We inspected the nodes for these trees and found that in each case the last node of the first tree and the first node of the second tree were in adjacent channels with an overlap fraction of 80\% and 75\% respectively, just below our threshold of 85\%. Given that their contours along the long axis of the filaments were nearly identical, combined with the greater thermal broadening in this region, we merged these tree groups into one filament each, leaving 7 trees in the region in total.

While we were unable to detect filaments that match our criteria in the lower resolution GBT data, we compare the \texttt{Filfinder} output on moment 0 maps integrated from 70-130\kms for both the GBT and GALFA-\ion{H}{i} data. These maps are shown in Figure \ref{fig:galfafilfinder}. For each panel of the Figure, we apply a less strict filtering process with a minimum filament aspect ratio of 1:2, minimum filament size of twice the beam size, and remove any filaments that touch the edge of the data. The looser criteria was needed in order to produce an output for the GBT data, which under these requirements returns one filament along the lower ridge of the Smith Cloud body. This filament is located in a region outside of the GALFA-\ion{H}{i} data. This comparison helps motivate why we were unable to find filaments in the GBT data following our standard requirements. By contrast, the GALFA-\ion{H}{i} data returns 70 structures (although artefacts are definitely still present), and there are cohesive filament structures at the locations of \texttt{fil3d} trees.

The different spatial resolutions also correspond to different width scales. For GALFA-\ion{H}{i} structures that \texttt{Filfinder} can compute the width of, i.e. excluding artefacts and other structures with obscure geometries, the average width is 32.9 pc at a distance of 12.4 kpc. By contrast, the computed width of the GBT structure in the lower ridge of the Smith Cloud body is 107.9 pc.

\subsection{Physical Properties}
Here we investigate various properties of the filament populations such as magnetic field alignment, column density, and temperature. The magnetic field alignment of filaments is shown in Figures \ref{fig:localbfield} and \ref{fig:bhist}. For Figure \ref{fig:localbfield} the color indicates $|\theta - \overline{\phi}|$ for each filament in degrees. The local filaments appear preferentially anti-aligned, with a median alignment (meaning the median difference between the orientation of a filament and its surrounding magnetic field) of 60\degr and a mean of 56\degr. The The anti-alignment is consistent across almost all local filaments, but seems more randomly oriented with respect to the \textit{Planck} data for the IVC and HVC populations. The median B-field alignment, as well as a summary of overall attributes for each group, are shown in Table \ref{tab:propertytable}. The mean computed polarization fraction within the merged mask area of the local filaments is .0725.

We make a temperature estimate for each filament by the following relation between gas line width and temperature:
\begin{equation}
    \label{tempeqn}
    \text{T} = 21.9~\text{K}\left(\frac{\Delta \text{v}}{\text{~km~s}^{-1}}\right)^2
\end{equation}

\begin{figure}
 \includegraphics[width=\columnwidth]{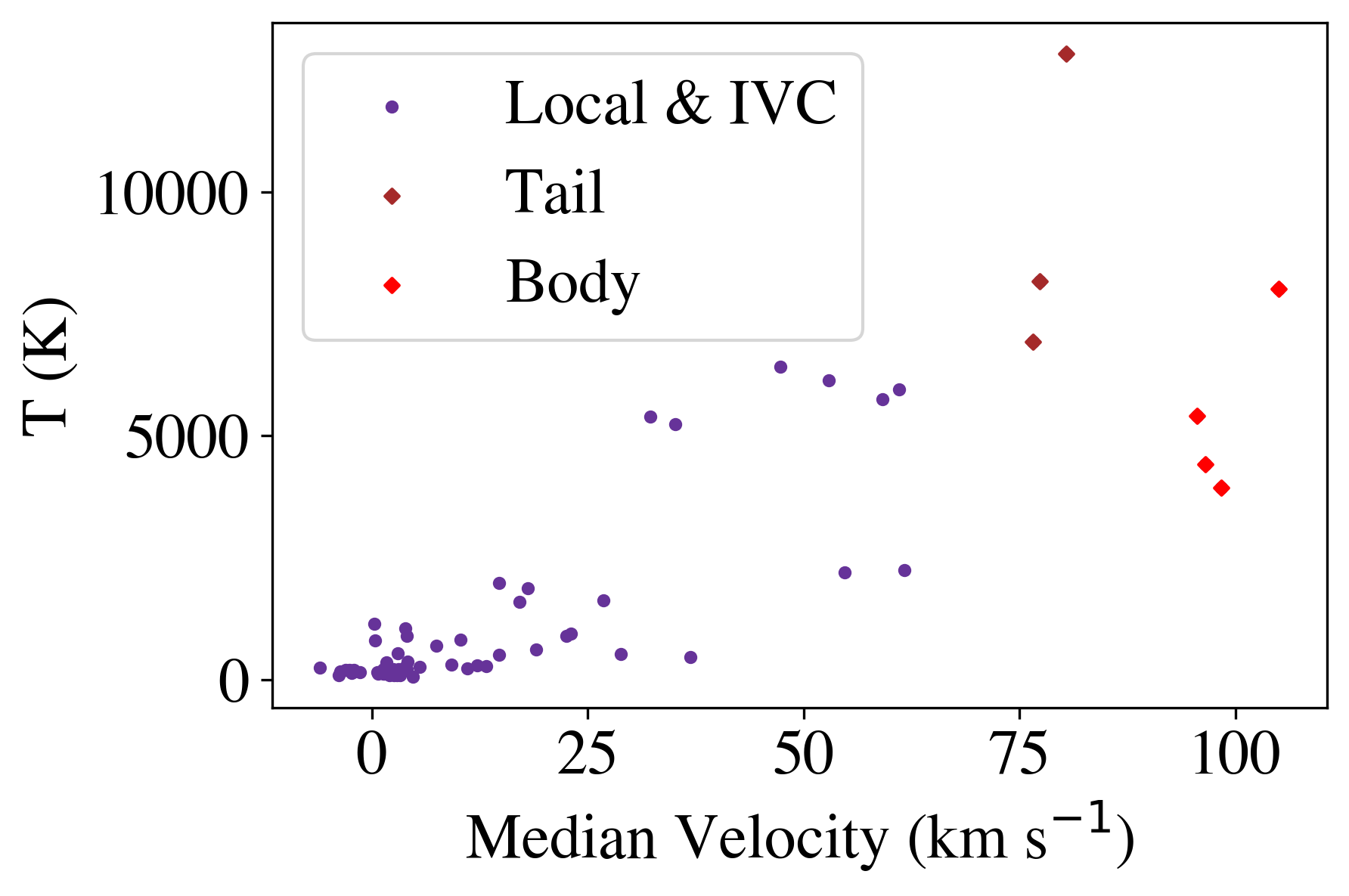}
 \caption{Filament temperature estimates as a function of median filament velocity. Points located in the tail of the Smith Cloud are maroon while points in the body are in orange. Filaments not in the Smith Cloud are purple.}
 \label{fig:fwhms}
\end{figure}
where $\Delta \text{v}$ is the FWHM as discussed in Section \ref{data}. Temperature as a function of median filament velocity is plotted in Figure \ref{fig:fwhms}. The median temperature estimate of the local filaments is 200 K, indicating they are part of the cold neutral medium (CNM). The median HVC population temperature estimate is $\sim$ 6900 K and the mean is $\sim$ 7100 K. This is expected since the Smith Cloud is known to have a warm neutral medium (WNM) and has been shown to have a \ion{H}{ii} mass comparable to its \ion{H}{i} mass \citep{Putman2003, Hill2009}. In particular, \cite{Hill2009} placed a lower bound temperature estimate on the ionized gas of $7000 \pm 2000$ K, which these results for the neutral gas are consistent with. Additionally, Figure \ref{fig:fwhms} shows a correlation between velocity and the temperature of the \ion{H}{i} gas along this line of sight. A least-squares linear regression of the data gives a best fit line of $T = 74.8 \cdot v_{LSR} + 154 \rm{K}$ with a correlation coefficient of $R=0.84$.

\begin{figure*}
 \includegraphics[width=0.99\textwidth]{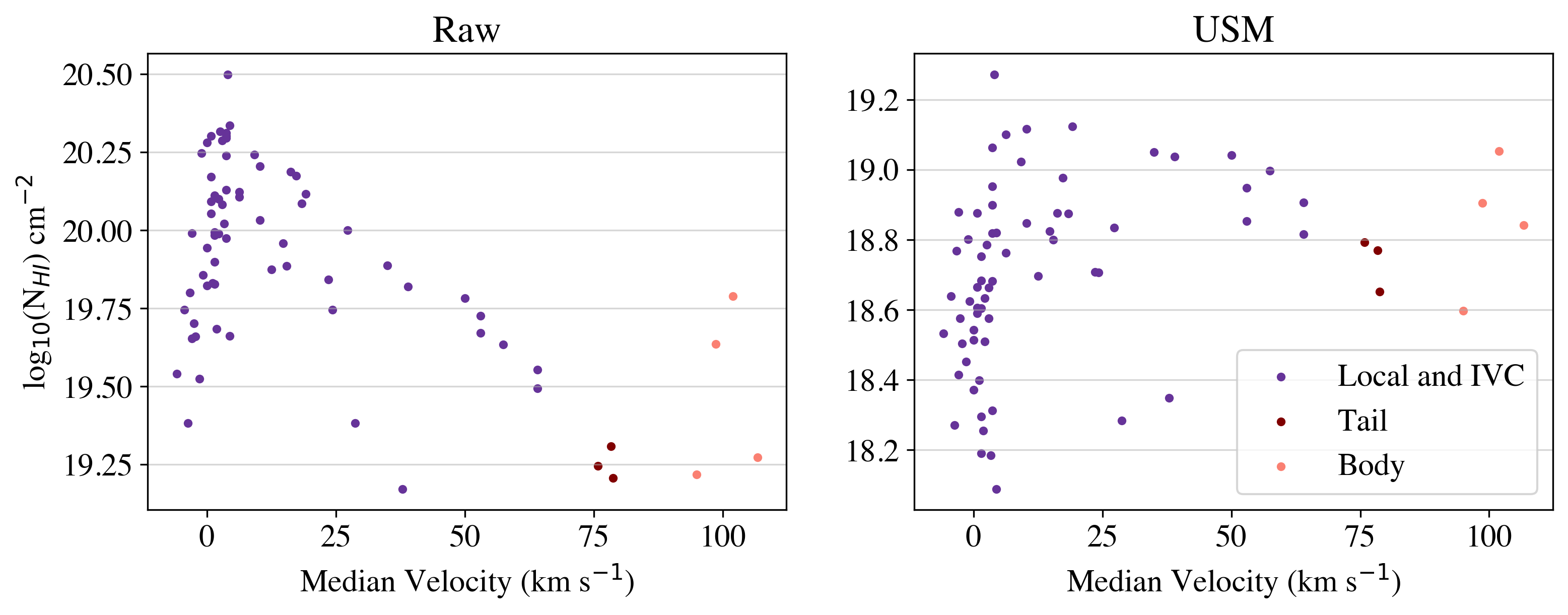}
 \caption{Median column density as a function of velocity for each filament generated from the raw (left) and USM (right) data. The local filaments around $\approx$ 0\kms are most greatly affected by the USM filter as the diffuse emission from the Galactic plane is removed, whereas the IVC and HVC filaments are less effected. The drop off in N$_\ion{H}{i}$ From 0 to 75\kms suggests a correlation between VLSR and z-height along the line of sight.}
 \label{fig:cdensity}
\end{figure*}

We also examine the column density of each filament. We evaluate the column densities by integrating over the FWHM of the line width fitted to each filament as discussed in Section \ref{data}. Figure \ref{fig:cdensity} shows the median column density within the merged mask area of each filament as a function of that filament's velocity for both the raw and USM data. The raw data shows a maximum density at v$_{LSR} \approx $ 0\kms\ followed by an exponentially decreasing relationship out to higher velocities. Since the USM data has had diffuse emission removed, particularly in the case of lower velocity filaments near the Galactic plane, this trend is not apparent in the USM data. More details on filament column densities, as well as filament spatial widths, are discussed in Putman et. al. (in prep.).

\begin{table}
 \caption{Properties for each filament group. The median value is reported for the B-field alignment, FWHM, and temperature.}
 \label{tab:propertytable}
 \begin{tabular}{lccccc}
  \hline
  Group & N & Velocity Range & $|\theta - \overline{\phi}|$ & FWHM & Temp.\\
   &  &(\kms) & (\degr) & (\kms) & (K) \\
  \hline
  Local & 52 & –8-20 & 60 & 3.1 & 200 \\
  IVC & 13 & 20-65 & 47 & 10.1 & 2200 \\
  HVC & 7 & 75-110 & 77 & 17.8 & 6900\\
  \hline
 \end{tabular}
\end{table}

\subsection{Smith Cloud Dust}

\begin{figure}
 \includegraphics[width=\columnwidth]{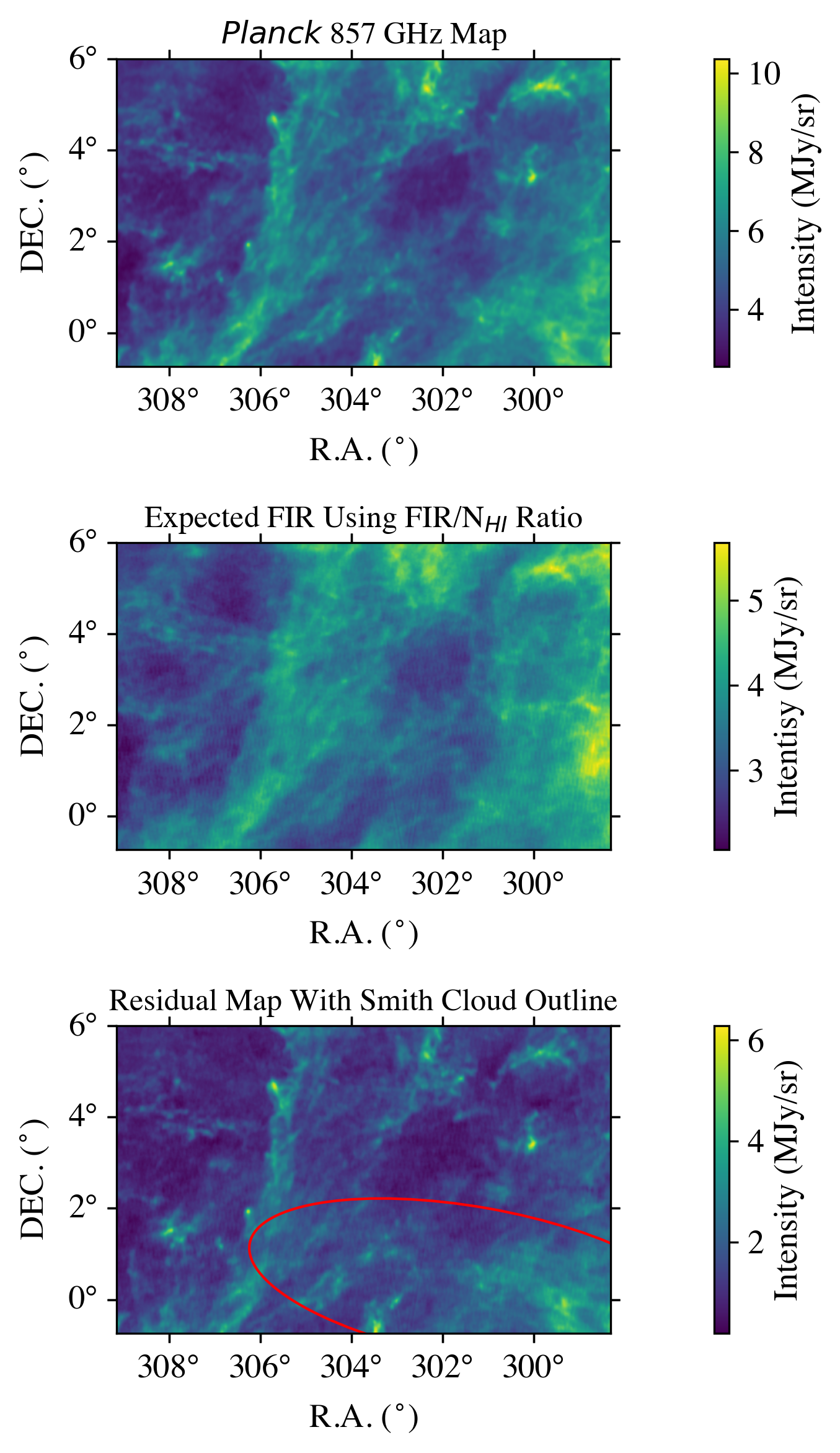}
 \caption{Summary of the investigation of a dust signature in the Smith Cloud. The top panel shows the \textit{Planck} 857 GHz map in the region of the Smith Cloud, which we treat as a proxy for FIR dust emission along the line of sight. The middle panel predicts the expected FIR emission as inferred from the Galactic \protect\ion{H}{i} by multiplying the FIR to \protect\ion{H}{i} ratio from \protect\cite{Planck2014dust}. Finally, we subtract the two maps to create the a residual map, shown in the bottom panel. The red outline indicates the position of the Smith Cloud and is the region within which we detect an excess compared to the rest of the region.}
 \label{fig:dustsample}
\end{figure}

We also attempted to investigate the dust content of the Smith Cloud. Due to the Smith Cloud's metallicity of half Solar \citep{Fox2016}, it is a good candidate to contain dust, with contrasting claims of detection appearing in the literature within the past year \citep{Minter2024, Vazquez2024}. If the Smith Cloud did contain detectable dust emission, then the \textit{Planck} far infrared (FIR) maps could in principle contain information on the Smith Cloud's plane-of-sky magnetic field orientation which could be compared to its filaments. To explore this idea further we sought out to see if there is evidence of an elevated FIR signal in the region of the Smith Cloud on the plane-of-sky in comparison to its surrounding region in our data cube. The summary of our investigation is shown in Figure \ref{fig:dustsample}. We begin by defining the region of the FIR map which we consider to be the region coinciding with the Smith Cloud as an ellipse with center $\alpha = 301\degr, \delta = 0.36\degr$, a semi-major and semi-minor axis of $5\degr$ and $1.67\degr$ respectively, and rotated $10\degr$ clockwise relative to the horizontal. The position and orientation of the ellipse is based on the corresponding position and orientation of the Smith Cloud in the \ion{H}{i} moment 0 map. We then created a map of the dust emission expected in this region from the Galactic emission alone by creating a column density map integrated from -60\kms < $v_{LSR}$ < 60\kms (which was chosen to ensure all local Galactic emission would be included but leave out HVCs along this line of sight) and multiplying by the $I_{857}/\text{N}_{\ion{H}{i}}$ ratio from \cite{Planck2014dust}. The $I_{857}$ map was chosen due to its similar native resolution to GALFA-\ion{H}{i}. We then subtracted the expected map from the \textit{Planck} 857 GHz map to create a residual map. After sampling 1000 random points both on and off the Smith Cloud region we do find a 99\% confidence level excess in the residual map of 0.4-0.55 MJy/sr (28\% - 38\% excess of the mean background value) which was confirmed using a 2-sample t-test. 

While this result is intriguing, there are far too many assumptions to claim with certainty this excess is directly from the Smith Cloud. For example, the residual map is positive over the majority of the map indicating the scaling ratio from \cite{Planck2014dust} was not a good fit in this region. This makes sense given the low Galactic latitude of this sight-line. However, we feel the excess detection in this region is worth noting and leave for future work to confirm whether or not the Smith Cloud is actually contributing to the FIR emission along this line of sight.

\section{Discussion}
\label{discussion}
In this paper we present the first quantification of velocity-resolved 3D \ion{H}{i} filaments in an HVC. We also identify foreground Galactic filaments with distinct velocity widths and orientations. Here we discuss the findings in the context of previous observational results and simulations.

\subsection{Local 3D Filaments}
Overall the local filaments are consistent with the 3D filaments detected by \cite{Kim2023} at high Galactic latitude in terms of line width and column density. The distinct exception is the filaments in this low Galactic latitude region are mostly anti-aligned with the plane-of-sky magnetic field. To further confirm this result, we compared our detected filaments to the \ion{H}{i} orientations in the Clark \& Hensley (C\&H) maps \citep{ClarkHensley2019}, which track the orientation of 2D coherent linear \ion{H}{i} structures using the Rolling Hough Transform. Using the same procedure that was done to compare alignment with the \textit{Planck} 353-GHz polarization map, we compute a median alignment between our 3D local filaments and the C\&H map of 15.7\degr, indicating a strong alignment which further validates the authenticity of our detected structures. We also checked our magnetic field results from \textit{Planck} by comparing it with the magnetic field derivation in \cite{Soler2016} and found our model matches their findings.

Filaments that are perpendicular to the Galactic magnetic field near the plane have previously been detected in \cite{Soler2020} and \cite{Soler2022}, which looked for \ion{H}{i} filaments at |b| < 1.25\degr and |b| < 10\degr, respectively. They associate regions of perpendicular alignment with direct supernovae feedback events as well as the cumulative effect from the Galactic fountain, where supernovae-heated gas rises above the Disc and flows radially outward \citep{ShapiroField1976,Bregman1980,Fraternali2006,Putman2012}. In either case, the dynamic feedback-driven environment likely dominates the orientation of the filaments.
In particular, \cite{Soler2022} finds more perpendicular filaments toward the inner galaxy, which can also be explained if caused by the Galactic fountain, since there is a greater concentration of supernovae remnants in the inner galaxy \citep[e.g.,][]{Green2015}. This is consistent with our findings, since the region studied here points toward the inner galaxy in the direction of the Sagittarius Arm. The local filaments in our sample are also preferentially oriented perpendicular to the Galactic plane.

Our local filament properties can be compared to \cite{Kim2023}, which observed a \ion{H}{i} filament population at high Galactic latitude. While we find a different behavior with respect to the magnetic field, we find a mean linewidth of 3.73$\pm$0.51\kms and median column density of $10^{20.04\pm0.06}$ cm$^{-2}$, compared to their findings of 3.1\kms\ and $10^{19.6}$ cm$^{-2}$, respectively. Our somewhat greater line-width is somewhat influenced by the fact that we employ a velocity resolution of 0.739\kms\ in the GALFA-\ion{H}{i} data whereas \cite{Kim2023} uses the narrower channels with a 0.184\kms resolution. 
Our slightly greater column density value makes sense as well in the context of our filaments being located closer to the Galactic plane. That being said, the column density of all of our filaments (Figure \ref{fig:cdensity}) is low compared to the general Galactic HI (above $10^{21}$ cm$^{-2}$).  Since the dust along this line of sight is likely associated with the densest HI, the HI filaments detected here may not be tracing exactly the same material as the \textit{Planck} data.  This could also explain the lack of alignment with the  Galactic magnetic field as traced by \textit{Planck}.

\subsection{HVC Filaments}
We find 7 filaments in the HVC known as the Smith Cloud, and they are oriented along its long axis. 
There is no singular reason for how filaments form, given their presence at varying scales and environments \citep{hacar23}. As the Smith Cloud approaches the Galactic disc, it encounters greater and greater pressure from the surrounding medium.  Stripping mechanisms ablate material from the cloud, and the stripped fragments are potentially condensed by the surrounding medium \citep{Heitsch2009}. It is already known that the head-tail structures of HVCs have more elongated lower-density structures in the tail \citep{putman11,for13}. Such conditions are more conducive to filament formation and may be consistent with warmer filaments towards the trailing tail of the HVC. 

The HVC magnetic field may also play a role in the filament formation. Previous work has shown local \ion{H}{i} filaments at high Galactic latitudes align with the plane of sky magnetic field as traced by dust polarization \citep[][]{Clark2014, Clark2015, Kim2023}, but to what extent the magnetic field effects the formation of the filaments is less certain. While we find the IVC and HVC filament alignment to be randomly oriented in regard to the magnetic field traced by 353 GHz dust polarization along the line of sight, the local ISM dominates the dust column with respect to the IVC dust \citep{Panopoulou2019}. Given the non-detection or limited detection of dust in HVCs \citep[e.g.,][]{peek09,Lenz2017}, it is reasonable to assume that the \textit{Planck} data does not capture IVC and HVC magnetic field information. Recent estimates of the strength of the Smith Cloud's magnetic field place it at 5 $\mu G$ \citep{Betti2019}, making the potential influence of the field substantial. Furthermore, \cite{Betti2019} created a model magnetic field for the Smith Cloud that would give a magnetic field orientation along the long axis of the Smith Cloud.  Since this is the same orientation as our detected Smith Cloud filaments, they are potentially aligned with its local magnetic field. 

The line widths of the Smith Cloud filaments are consistent with them being part of a warm neutral medium within the cloud.   It will be interesting to explore the distribution of filaments and filament temperatures further with future high resolution HI observations, such as ASKAP \citep{ASKAP2020} and FAST \citep{FAST2024}.  Comparing Figures \ref{fig:upperfilaments} and \ref{fig:galfafilfinder}, we see that the GBT moment 0 map identifies material in the lower midsection, as well as the lower extending ridge.  With higher-resolution data in the region not covered by GALFA-\ion{H}{i}, \texttt{fil3d} would likely return filaments in these areas as well.

\subsection{Overall Properties}

The trends in velocity FWHM and column density in Figures \ref{fig:fwhms} and \ref{fig:cdensity} can be understood from the perspective that the farther out we look along this line of sight (represented at some level by the filament velocity) the greater the $|z|$ value from the plane becomes. With the FWHMs (and temperatures derived from then using Equation \ref{tempeqn}), we see the local region has line widths consistent with the CNM but then scales to the WNM as $|z|$ increases as we reach the Smith Cloud, which sits 3 kpc below the disc. For column densities, we see the emission in the raw data peak around 0\kms and drop off as we go further from the plane, which is consistent with the overall observed \ion{H}{i} distribution of the Galactic ISM \citep{Kalberla2009}.

The separate filament groups also correspond to different physical scales. Assuming a distance of 200 pc, with the given GALFA-\ion{H}{i} spatial resolution the local filaments have an average width (defined as twice the radial profile of the root node computed in \texttt{Filfinder}) of 0.66 pc, whereas at a distance of 12.4 kpc the HVC filaments have a mean width of 38 pc. These filaments are significantly larger than the sub-pc local ISM filaments and have not been studied as extensively in the literature. As such, their formation mechanism may vary from the smaller scale filaments and there may be multiple filaments embedded in the large linear structures. As a precaution, we note that the computed widths seem to scale with both the resolution of the data as well as the USM kernel size (Putman et al. in prep). Therefore, these value should be treated as upper limits on the widths given our resolution.

\section{Conclusion}
\label{conclusion}

In this work, we report the presence of 3D velocity resolved filaments in the Smith Cloud using our filament detection algorithm \texttt{fil3d}. They are physically aligned along the long axis of the cloud. The filaments' temperatures and column densities are consistent with what would be expected for an HVC in the warm neutral medium infalling towards the Galactic disc. We compare the HVC filaments to local filaments at a low Galactic latitude along the line of sight and find that neither population are aligned with the plane-of-sky magnetic field orientation inferred from Faraday rotation. We conclude as a result that dynamic processes in the disc likely dominate the local filament orientations.

Future work entails searching for filaments in other HVC complexes to see if properties are shared across different objects. 
Complex C, another well-studied HVC, has a well-constrained distance \citep{Thom2008} and is at least partially covered by GALFA-\ion{H}{i}'s observation window \citep{hsu11}. Large area surveys being completed with ASKAP and FAST will provide additional maps of HVCs at high enough resolution to search for 3D filaments \citep[e.g.][]{Ma2023}. For the Smith Cloud, in particular, a full high-resolution map is needed to study the filamentary structure of the entire cloud.  An increased knowledge of the magnetic field of the Smith Cloud, as well as halo clouds in general, is also required to test if the filaments align with the cloud's magnetic field. It is also possible that filaments may be observed in ionized hydrogen as well. The Smith Cloud has already been mapped in H-$\alpha$ \citep{Hill2009} at a low resolution, but doing so at higher resolution may reveal a network of H-$\alpha$ filaments as seen in Galactic ISM H-$\alpha$ surveys \citep[e.g.][]{mdwdr0}.

In terms of \texttt{fil3d}, we acknowledge that it is not automated in its present form and fine tuning of the \texttt{FilFinder} parameters is required for a given region and dataset to maximize the results. Section \ref{fil3dparams} serves as a basis for tuning the parameters to other regions which we hope can prove useful when applying the program to a new region for the first time. Nevertheless, we feel that \texttt{fil3d} in its present form has broad applicability to characterizing cohesive 3D-linear structures in the ISM. This will play a very important role into understanding how \ion{H}{i} ISM filaments are formed, in what Galactic environments they exist in, as well as characterizing the properties and distribution of resolved three dimensional structures in the ionized Galactic gas.

\section*{Acknowledgements}

We thank the anonymous referee whose feedback improved this manuscript. The authors thank Susan Clark for many helpful discussions and Jay Lockman for sharing the GBT data of the Smith Cloud. The authors also acknowledge useful discussions with Julia Homa, Noor Aftab, and Andrew Zhang.
This project makes use of NumPy \citep{numpy}, AstroPy \citep{Astropy2013, Astropy2018}, Matplotlib \citep{matplotlib2007}, \textit{Filfinder} \cite{Koch2015}, and Healpy \citep{healpy}

%%%%%%%%%%%%%%%%%%%%%%%%%%%%%%%%%%%%%%%%%%%%%%%%%%
\section*{Data Availability}

This paper makes use of data from the Galactic Arecibo L Band Feed Array (GALFA) \ion{H}{i} Data Release 2 which are publicly available to download at https://purcell.ssl.berkeley.edu/, the \ion{H}{i} 4$\pi$ Survey, available at http://cade.irap.omp.eu/dokuwiki/doku.php?id=hi4pi, as well as observations from the \textit{Planck} Collaboration, which are publicly accessible at http://www.esa.int/Planck. In addition, this paper made use of Smith Cloud data from the Robert C. Byrd Green Bank Telescope, Green Bank, WVA, which is available upon reasonable request to the authors. \texttt{fil3d} is publicly available to use and can be downloaded at https://github.com/LLi1996/fil3d. Data and code corresponding to the plots in this paper are also available upon reasonable request to the authors.



\bibliographystyle{mnras}
\bibliography{main} % if your bibtex file is called example.bib

%\begin{thebibliography}{99}
%\end{thebibliography}

%%%%%%%%%%%%%%%%%%%%%%%%%%%%%%%%%%%%%%%%%%%%%%%%%%

%%%%%%%%%%%%%%%%% APPENDICES %%%%%%%%%%%%%%%%%%%%%

%%%%%%%%%%%%%%%%%%%%%%%%%%%%%%%%%%%%%%%%%%%%%%%%%%

% Don't change these lines
\bsp	% typesetting comment
\label{lastpage}
\end{document}